# On Energy-Efficient Time Synchronization based on Source Clock Frequency Recovery in Wireless Sensor Networks

Kyeong Soo Kim, Sanghyuk Lee, and Eng Gee Lim
Department of Electrical and Electronic Engineering, Xi'an Jiaotong-Liverpool University, Suzhou, P. R. China
{Kyeongsoo.Kim, Sanghyuk.Lee, Enggee.Lim}@xjtlu.edu.cn

*Abstract*—In this paper we study energy-efficient time synchronization schemes with focus on asymmetric wireless sensor networks, where a head node, which is connected to both wired & wireless networks, is equipped with a powerful processor and supplied power from outlet, and sensor nodes, which are connected only through wireless channels, are limited in processing and battery-powered. It is this *asymmetry* that we focus our study on; unlike existing schemes saving the power of all sensor nodes in the network (including the head node), we concentrate on battery-powered sensor nodes in minimizing energy consumption for synchronization. Specifically, we discuss a time synchronization scheme based on source clock frequency recovery, where we minimize the number of message transmissions from sensor nodes to the head node, and its extension to network-wide, multi-hop synchronization through gateway nodes.

**Keywords-Time synchronization; source clock frequency recovery; packet delay; wireless sensor networks**

## I. INTRODUCTION

Real-time wireless data acquisition networks, e.g., large-scale wireless sensor networks (WSNs) deployed over a vast geographical area, have been the focus of extensive studies due to their versatility and broad range of applications. Time synchronization is one of critical components in WSN operation, as it provides a common time frame among different nodes. It supports functions such as fusing data from different sensor nodes, time-based channel sharing and media access control (MAC) protocols, and coordinated sleep wake-up node scheduling mechanisms [1]. As a sensor node is a low-complexity, battery-powered device, energy efficiency is the key in designing schemes and protocols for WSNs.

In a typical WSN, a master/head node is equipped with a powerful processor, connected to both wired & wireless networks, and supplied power from outlet because it serves as a gateway between the WSN & a backbone and a center for fusion of sensory data from sensor nodes, which are limited in processing and electrical power because they are connected only through wireless channels and battery-powered. It is this *asymmetry* that we focus our study on; unlike existing schemes which save the power of all WSN nodes including the head (e.g., [2] and [3]), we concentrate on battery-powered sensor nodes, which are many in numbers, in minimizing energy consumption for synchronization. Specifically, in this paper we discuss a time synchronization scheme based on the source clock frequency recovery (SCFR) [4], where we minimize the number of message transmissions at sensor nodes because the energy for packet transmission is typically higher than that for packet reception [5]. We also discuss its extension to network-wide, multi-hop synchronization through gateway nodes.

## II. SCFR-BASED WSN TIME SYNCHRONIZATION

The major idea is to allow independent, unsynchronized slave clocks at sensor nodes but running at the same frequency as the master clock at a head node through the asynchronous SCFR schemes described in [4], which need only reception of messages with timestamps, while carrying out the two-way message exchange, which is unavoidable for recovery of clock offset in existence of propagation delay [6], using normal data packets to reduce the number of transmissions at sensor nodes. In this way, the head node can estimate time offsets of sensor nodes and correctly interpret the occurrence of data measurements with respect to its own master clock.

Fig. 1 illustrates this idea in comparison to ordinary schemes shown in Fig. 1 (a) that are based on two-way message exchange: First, the proposed scheme shown in Fig. 1 (b) does not have periodic, dedicated two-way message exchange sessions with special control messages like "Request" and "Response"; instead, the two-way message exchange is carried out using an ordinary message from a sensor node and the most recent message from the head, both of which have embedded timestamps. Secondly, the direction of two-way message exchange in the proposed scheme is reversed, i.e., it is the master that requests, not the slave, unlike the existing schemes; as a result, the master knows the current status of the slave clock, but the slave does not. So the information of slave clocks (i.e., time offsets with respect to the master clock) is centrally managed at the head node.

Note that, for operations like coordinated sleep wake-up node scheduling, the head node first adjusts the time for future operation (with respect to its own master clock) based on the time offset of a recipient sensor node before sending it to that node in the proposed scheme. In this way, even though sensor nodes in the network have clocks with different time offsets, their operations can be coordinated based on the common master clock in the head node.

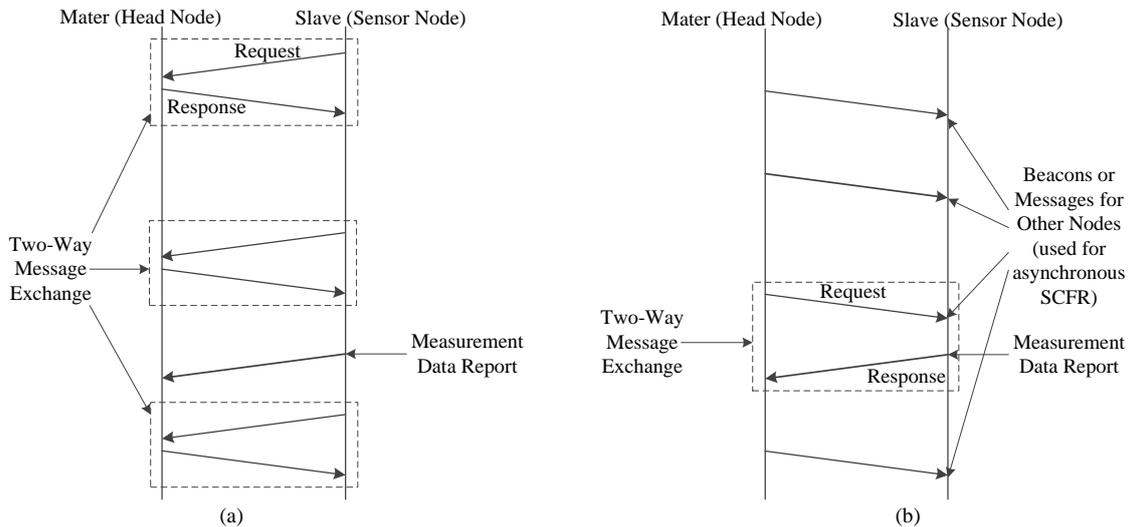

Figure 1. Reducing message transmissions at sensor nodes: (a) A scheme based on two-way message exchange as in time-sync protocol for sensor networks (TPSN) [7] and (b) the proposed scheme.

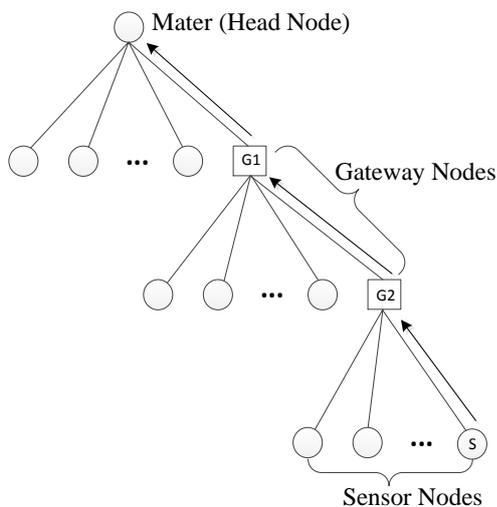

Figure 2. Extension of the proposed time synchronization scheme to a hierarchical structure for network-wide, multi-hop synchronization through gateway nodes.

Fig. 2 shows how the proposed scheme can be extended to a hierarchical structure for network-wide, multi-hop synchronization through gateway nodes which act as both head nodes (for nodes below) and normal sensor nodes (for nodes above): For example, consider the message transmission from the sensor node S to the head node through the two gateway nodes G1 and G2 as shown in Fig. 2. Because G2 acts as a head node for the sensor node S, it translates the value of time stamp based on the information on the time offset of S. Then G2 relays the message from S to G1 with translated time stamp value (with respect to its own clock). From G1's point of view, by the way, G2 is just one of sensor nodes it manages. Again, based on the information on time offset of G2, G1 translates the value of time offset with respect to its own clock and relays the message to the head node. Finally, the head nodes receives the message from S, which is just relayed by G1, and translates the time stamp value based on the information on the time offset of G1 it manages. In this way, the head node can obtain the event & related data and its occurrence in time reported by S with respect to its own master clock.

III. SUMMARY

In this paper we have proposed an energy-efficient time synchronization scheme for asymmetric WSNs, which is based on the asynchronous SCFR and master-initiated two-way message exchange to minimize the number of message transmissions at sensor nodes. We have also shown how the proposed scheme can be extended to a hierarchical structure for network-wide, multi-hop synchronization through gateway nodes.

ACKNOWLEDGMENT

This work was supported by the Centre for Smart Grid and Information Convergence (CeSGIC) at Xi'an Jiaotong-Liverpool University.